\documentclass[12pt,pdf]{article} 
\usepackage{graphicx,amssymb,amsfonts,amsmath,bm,url,color,adjustbox} 
\usepackage{arydshln,hyperref}

\newtheorem{remark}{Remark}

\newtheorem{example}{Example}
\def\pmatrix{\left(\begin{array}}
\def\endpmatrix{\end{array}\right)}

\def\PapRank{{\it Paper\-Rank}}
\def\AutRank{{\it Author\-Rank}}
\def\NumAut{{\it \#Auth}}
\def\NumRef{{\it \#Ref}}
\def\NP{\#Pub}
\def\NC{\#Cit}
\def\SPR{$\sum$P.R.}
\def\SC{$\sum$Cit}

\title{Implementation of the  {\em PaperRank} and {\em AuthorRank} indices in the Scopus database}
\author{Pierluigi Amodio\footnote{Dipartimento di Matematica, Universit\`a di Bari, Via Orabona 4, 70125 Bari, Italy.\qquad\qquad e-mail:\,{\tt pierluigi.amodio@uniba.it}}
\and Luigi Brugnano\footnote{Dipartimento di Matematica e Informatica ``U.\,Dini'', Universit\`a di Firenze, Italy.\qquad\qquad e-mail:\,{\tt luigi.brugnano@unifi.it}} \and Filippo Scarselli\footnote{e-mail:\,{\tt f.scarselli@tiscali.it}}}

\begin{document}

\maketitle

\begin{abstract}
We implement the \PapRank\ and \AutRank\ indices introduced in [Amodio \& Brugnano, 2014] in the Scopus database, in order to highlight quantitative and qualitative information that the bare number of citations and/or the h-index of an author are unable to provide. In addition to this, the new indices can be cheaply updated in Scopus, since this has  a cost comparable to that of updating the number of citations. Some examples are reported to provide insight in their potentialities, as well as possible extensions.


\bigskip

\end{abstract}

\section{Introduction}
This paper deals with the interesting question of ranking authors and scientific publications. This is a very important and debated topic, especially since the coverage of bibliometric databases has made it possible to correctly analyze the scientific path of each author, at least for those of the last 50 years \cite{DoMe15,GaHiAlFa18,YaDiSu11}.
Today, citation analysis represents a reliable formal tool for quantitative scientific assessment.

For our research we use the Scopus database \cite{Scopus} which assigns to each author an identifier (the Scopus A.Id ) in order to count, on a Scopus dedicated page, the list of the authored papers and the references therein. This allows us to immediately know, not only the number of papers, but also the number of citations of each author in the database. To these values Scopus adds, thanks to its simplicity of calculation, the h-index \cite{Hi05}, that is, a value equal to $k$ if the author authored $k$ papers with at least $k$ citations.

These parameters, although widely used, have drawbacks which lead to results that are sometimes not meaningful (see, e.g., \cite{BrNa21,KrKoNg21,MoNyMoAnSa21} published this year (2021) and Section \ref{SecRes}). For this reason, many other indices have been proposed (see, for example, \cite{BVdSHC09}). Some of them (but this is only a very short list) just follows from the h-index: m-index \cite{Hi05}, g-index \cite{Eg06,Eg06b}, e-index \cite{Zh09}, h$^{\text c}$-index \cite{SiKaMa07}. Others, as the i10-index (used by Google Scholar \cite{GScholar}) which counts the number of papers with more than 10 citations, derives from a different observation of the data. Anyway, none of these indices takes into account the fact that, for example: 
\begin{itemize}
\item papers from different subject areas usually use bibliographies with quite different lengths;
\item within the same subject area, the number of citations increases over time;
\item review papers have very long reference lists while very targeted research works usually have a small bibliography;
\item self citations may be a significant part of the citations.
\end{itemize}

For this reason other indices have been introduced which, however, require a more complex processing of the data contained in the database. In this regard, an algorithm, that can certainly be taken into consideration as a reference, is the {\it PageRank} of Google \cite{PaBr98} which, when a Google search is performed, assigns a score to each page and then provides the user with a ranking of the most interesting pages \cite{Google}. The {\it PageRank} algorithm is based on the number of incoming links and the weighting of the linking page. Irrespective of the content of a given page, this latter is better evaluated when important pages link to it. The ranking of a site is therefore recursive to the assessment of the linking page (the {\it PageRank} depends on the number of pages linking to it, and the quality of these pages).

Bringing this idea back to our problem, the idea is that the importance of a paper depends not only on the citations received, but also on the importance of the papers that make these citations. Several authors have exploited this idea to define new indices, for example P-Rank \cite{YaDiSu11}, AuthorRank \cite{LiBoNeVa05}, Article Rank \cite{LW09}, Y-factor \cite{BoRoVa06}, CiteRank \cite{WaXiYaMa07} and Eigenfactor \cite{Be07,BeWe10}. The problem is that the calculation of these indices requires a high computational cost, and an update of the reference database (which happens at least weekly on Scopus) requires the complete recalculation of the computed indices. For this reason, the numerical tests carried out with these indices, always refer to much smaller databases, or analyze journals of a given subject area \cite{ChXiMaRe06,Fiala,MaGuZh07}.

In \cite{AmBr14}, a mathematically correct simplification of the {\it PageRank} model has been proposed, which requires a computational cost comparable to that of the h-index. For this reason, it could be easily implemented in the Scopus database. A further feature of this new index is that it is additive, and still preserving the main properties of the Google Rank.  This means that it can be easily updated, by taking into account possible new citations.

With these premises, the structure of the paper is as follows: in Section~\ref{SecRank} we recall the main facts about the \AutRank\ and \PapRank\ presented in \cite{AmBr14}; in Section~\ref{SecWeb} we introduce a web page that has been set up for computing these new parameters by means of a few Scopus API keys \cite{apik}, and in Section~\ref{SecRes} we provide some simple examples that highlight their potential; finally, in Section~\ref{SecNewind} some new indices are defined, that could be derived (for free) to provide further insight.

\section{Materials and methods: definition of \PapRank\ and \AutRank } \label{SecRank}

The idea behind the definition of \PapRank\ in \cite{AmBr14} derives, as mentioned above, from that of the Google {\it PageRank} \cite{PaBr98} which, in our case can be rewritten and summarized in this way: \\
\centerline{ \em ``an important paper is cited by important papers’'.}

From a mathematical point of view, starting from a database containing the papers, each with its own references, it is possible to obtain the importance of a paper by calculating the eigenvector $v$ associated with the unit eigenvalue of the matrix $S=LF^{-1}$, where $L$ is the citation matrix (of size equal to the size of the database)
\begin{equation}\label{L}
L = (\ell_{ij}), \qquad
\ell_{ij} = \left\{ \begin{array}{cl} 1 &\mbox{~if paper $j$ cites paper $i$,}\\ 0 &\mbox{~otherwise,}\end{array}\right.
\end{equation}
and $F$ is a diagonal matrix, whose diagonal entries are defined as
\begin{equation}\label{F}
f_j = \sum_i\ell_{ij}.
\end{equation}
We observe that $f_j$ is nothing but the number of references in the bibliography of the paper $j$.
For the sake of simplicity, we assume that $f_j>0$, even though this will be not strictly necessary, as explained below  (see \cite{AmBr14} for more details).
By setting, hereafter, $e$ the vector of length $n$ with all entries equal to one, one easily realizes that
$$e^\top S = e^\top L F^{-1} = \left(f_1,\,\dots,f_n\right)^\top F^{-1} = e^\top,$$
so that  $\lambda=1$ is an eigenvalue of matrix $S$. Similarly, one has that $\|S\|_1=1$, where we recall that the 1-norm of a matrix is the maximum among the sums of the absolute values of the entries on each row (in particular, the sum on each row of $S$, which is a non-negative matrix, is 1). Consequently , since the spectral radius of $S$ is less than or equal any natural norm, the unit eigenvalue is the one of maximum modulus.

The calculation of the eigenvector $v$ requires an information of global nature, as in the case of Google {\it PageRank} (see, for example, \cite{KaHaGo04,KaHaMaGo03} where efficient strategies are implemented to reduce the corresponding computational cost). In addition, any update of the database would require a complete recalculation of the vector $v$. For this reason, in \cite{AmBr14} the following, much cheaper, approximation was proposed, 
\begin{equation}\label{v0}
v_0:= Se = LF^{-1} e. 
\end{equation}
We observe that the vector $v_0$ can be also regarded as the first step of the power method, applied for computing the right eigenvector of the unit eigenvalue of $S$, starting from $e$.

This is the best approximation we can derive by using only a local information: in fact, its computational cost equals that for computing the bare number of citations (i.e., the vector $Le$) and turns out to be additive, that is, an update of the database requires to update the approximation {\em only taking into account} the new entries (i.e., the newly added citations). 

Hereafter, we shall refer to the vector $v_0$ as the \PapRank\ \cite{AmBr14}. Moreover:
\begin{itemize}
\item  we shall denote by $n$ the number of papers in the database (i.e., matrix $L$ is of size $n\times n$);
\item for each paper $P_j$ in the database, we set:
\begin{itemize}
\item[$>$] $\NumAut(P_j)$ the number of authors of the paper,
\item[$>$] $\NumRef(P_j)$ the number of references in its bibliography (i.e., $f_j$),
\item[$>$] $n_j$ the number of papers citing it.
\end{itemize}
\end{itemize}
Let $c_{1,j},\dots c_{n_j,j}$ be the indices of the papers citing paper $P_j$. Then, according to (\ref{L})--(\ref{v0}), we define:
\begin{equation}\label{Prank}
\PapRank(P_j) =  \sum_{i=1}^{n_j} \frac{1}{\NumRef(P_{c_{i,j}})}, \qquad j=1,\dots,n. 
\end{equation} 

\noindent From this definition we also get that of \AutRank. In more detail: 
\begin{itemize}
\item let $m$ be the number of all authors of all papers in the database;
\item for each author $A_k$, we set $m_k$ the number of her/his papers.
\end{itemize}
Let $b_{1,k},\dots b_{m_k,k}$ be the indices of the papers written by the author $A_k$. Then, we set:
\begin{equation}\label{Arank}
\AutRank(A_k) =  \sum_{j=1}^{m_k} \frac{\PapRank(P_{b_{j,k}})}{\NumAut(P_{b_{j,k}})}, \qquad k=1,\dots m. 
\end{equation}

\begin{remark}\label{dispro}
 In other words, the \PapRank\ of each paper is equally divided among all its authors. Therefore, it is easy to show that (see Theorem 2 in \cite{AmBr14}):
$$ \sum_{j=1}^n \PapRank(P_j) = \sum_{k=1}^m \AutRank(A_k). $$
It is worth mentioning, however, that any discrete probability distribution could in principle be used (taking into account, as an example, the order of the authors), even though we shall not consider this possibility, hereafter.
\end{remark}

\begin{example} To better explain how the \PapRank\ and the \AutRank\ can be easily updated, let us assume that a given author, having  \AutRank\,=\,10, receives a new citation concerning one of her/his paper, having \PapRank\,=\,2, written together with another co-author, and that this citation is done by a paper with 5 references. As a result:
\begin{itemize}
\item the \AutRank\ is updated as\quad {\rm new-\AutRank~=~10\,+\,$\frac{1}{2\times 5}$~=~10.1};

\item the \PapRank\ is updated as\quad {\rm new-\PapRank~=~2\,+\,$\frac{1}{5}$~=~2.2};
\end{itemize}
thus confirming the  incremental nature of such indices.
\end{example}

\section{A web page to compute the \AutRank} \label{SecWeb}

We have developed a web page which is able to query the Scopus database in order to compute the rank of each paper (\PapRank) and, consequently, of each author (\AutRank). This is possible since in Scopus:
\begin{itemize}
\item any author is identified by a unique numeric code (i.e., the Scopus Author Identifier mentioned above, A.Id  in short);
\item any paper is identified by an analogous alphanumeric string (Electronic Identifier, eid in short), beginning with the characters {\tt 2-s2.0-}.
\end{itemize}

Scopus, beyond immediately providing some information about each author (including number of publications and h-index) and publication (for example the bibliography and the number of citations), allows to formulate additional queries through API keys that are well explained starting from the web page \cite{apik}.

Through these API keys it is possible to retrieve the information needed to calculate the \PapRank\ and the \AutRank. In particular, to compute the \PapRank\ of a given paper:
\begin{itemize}
\item the number of the received citations;
\item for each citation, the length of the bibliography of the citing paper;
\item the \PapRank\ is then computed according to (\ref{Prank}).
\end{itemize}
Obviously the first value is the only one that can be updated (increased) over time.
According to what has been exposed in the previous section, to compute the \AutRank\ of a specific author, one has to compute the number of the author's publications and, for each publication, the corresponding \PapRank, as specified above. Then the \AutRank\ is computed according to (\ref{Arank}). 

Therefore, the computation of the \PapRank\ requires a number of queries equal to the number of citations of the considered paper.  Correspondingly, the computation of the \AutRank\ requires a number of queries equal to the number of  papers of the authors plus the sum of the queries to compute the \PapRank\ of each one of such papers. With the obtained informations, both the \PapRank\ and the \AutRank\ are calculated by means of simple weighted averages, according to (\ref{Prank}) and (\ref{Arank}). Furthermore, since these operations are associative, we stress that updating the \AutRank\ and/or the \PapRank\   only require to consider:
\begin{itemize}
\item possible new citations of a given paper;
\item possible new papers of a given author.
\end{itemize}

Actually, the web page allows to find an author starting from his name and/or the A.Id and compute the \AutRank\ and the \PapRank\ of each document to date. Moreover, since all the information allowing the calculation is stored (with the data of computation), it is also possible to dynamically update these indices.

\begin{remark}\label{add1}
We would like to emphasize that,  should the value of the \PapRank\ be added to the Scopus record of each paper, then each new citation of that paper would result in an increase in its \PapRank\  equal to 1 over the length of the bibliography of the paper making the new citation.

Equivalently, any newly added paper in the Scopus database would increase the \PapRank\ of each paper referenced in its bibliography by  1 over the length of the bibliography itself (in fact, the sum of these updates has to be one).

Similarly, any update in the \PapRank\ of a given paper, would result in an update of the \AutRank\ of its authors equal to the increase of the \PapRank\ divided by the number of the authors.\footnote{We want to emphasize that the web page has been set up for exhibition purposes, and relies on the use of the API keys, which Elsevier has granted us for this research.}
\end{remark}

\section{Results} \label{SecRes}

The aim of this section is to highlight the potential of the \PapRank\ and the \AutRank. For this purpose, we shall denote:
\begin{enumerate}
\item \NC\  the number of citations obtained by a given paper; 
\item  \NP\ the number of publications of a given author;
\item \SC\ the sum of citations received by (all papers of) a given author;
\item \SPR\ the sum of the \PapRank\ of (all papers of) a given author.
\end{enumerate}
The first three parameters are well highlighted in the Scopus database, for any paper/author. However, they may not be sufficient to properly rank them. In particular, the following aspects can be stressed: 
\begin{description}
\item[a)] Authors having a low h-index may have a high \AutRank. 
\item[b)] A paper with a higher number of citations than another one, may have a lower \PapRank.
\end{description} 

\noindent Concerning point {\bf a)}, we report the following two examples:

\begin{itemize}
\item the authors of the seminal paper \cite{PaBr98}, Brin and Page, have a low h-index. In fact, Brin's h-index is 10. Moreover, Page has two different  A.Id in Scopus, one with h-index 3, and one with h-index 1. We  computed the corresponding \AutRank\ and, in particular, that of Page is nothing but the bare sum of the \AutRank\ of his two profiles. As a result, we obtained
\AutRank\ 221.76 and 166.36 for the two authors, respectively, largely due to the citations of their most important paper \cite{PaBr98} and its reprint \cite{BrPa12} (having \PapRank\ 319.37 and 17.05, respectively); 

\item the two 2013 Nobel Prizes in Physics, F.\,Englert and P.W.\,Higgs  have a very different h-index, 27 the former and 9 the latter. Nevertheless, \SPR\ of their respective papers (at least, those contained in Scopus) turn out to be comparable (161.31 vs. 182.345), and the \AutRank\ are reversed (73.28 vs. 182.24), due to the fact that almost all the papers by Higgs are with only one author.
\end{itemize}

\noindent Concerning point {\bf b)}, we observe that this naturally happens for papers from different subject areas. Nevertheless, this may happen also for papers within the same subject area and by the \underline{same author}. For example, analyzing the papers with the highest number of citations in the huge bibliography of Karplus ($\AutRank = 1010.29$, see Table \ref{Tab2}) it turns out that a work published in 2009 with 34 co-authors has over 4500 citations and a $\PapRank = 78.11$, while a work (with a single name) published in 1958 with 2600 citations has a $\PapRank = 91.14$. This is due to the fact that, in recent years, as the h-index has become more and more important for academic evaluations, the bibliography length has grown a lot. Consequently, a citation today generally weighs much less than one obtained several years ago.

To better highlight the features of the new indices, it would be appropriate to compare authors having the same \SC\ but, unfortunately, Scopus, through the Author Search, allows to sort (apart from the useless, for our purpose, alphabetical order) only by the h-index and by the Document Count. For this reason, we now perform some test concerning authors with the same h-index. In particular, Table\,\ref{Tab1} contains the 31 authors whose name begins with the letters ``AL'' and having, on 15 January 2021, an h-index = 48. These authors (see Table\,\ref{Tab1}) come from 12 different ``Subject Areas'' (Scopus identifies 26 of them) which, for the sake of completeness, are fully explained in Table\,\ref{Tab3}. 

\begin{figure}[htb]
\begin{center}
\includegraphics[width=\textwidth]{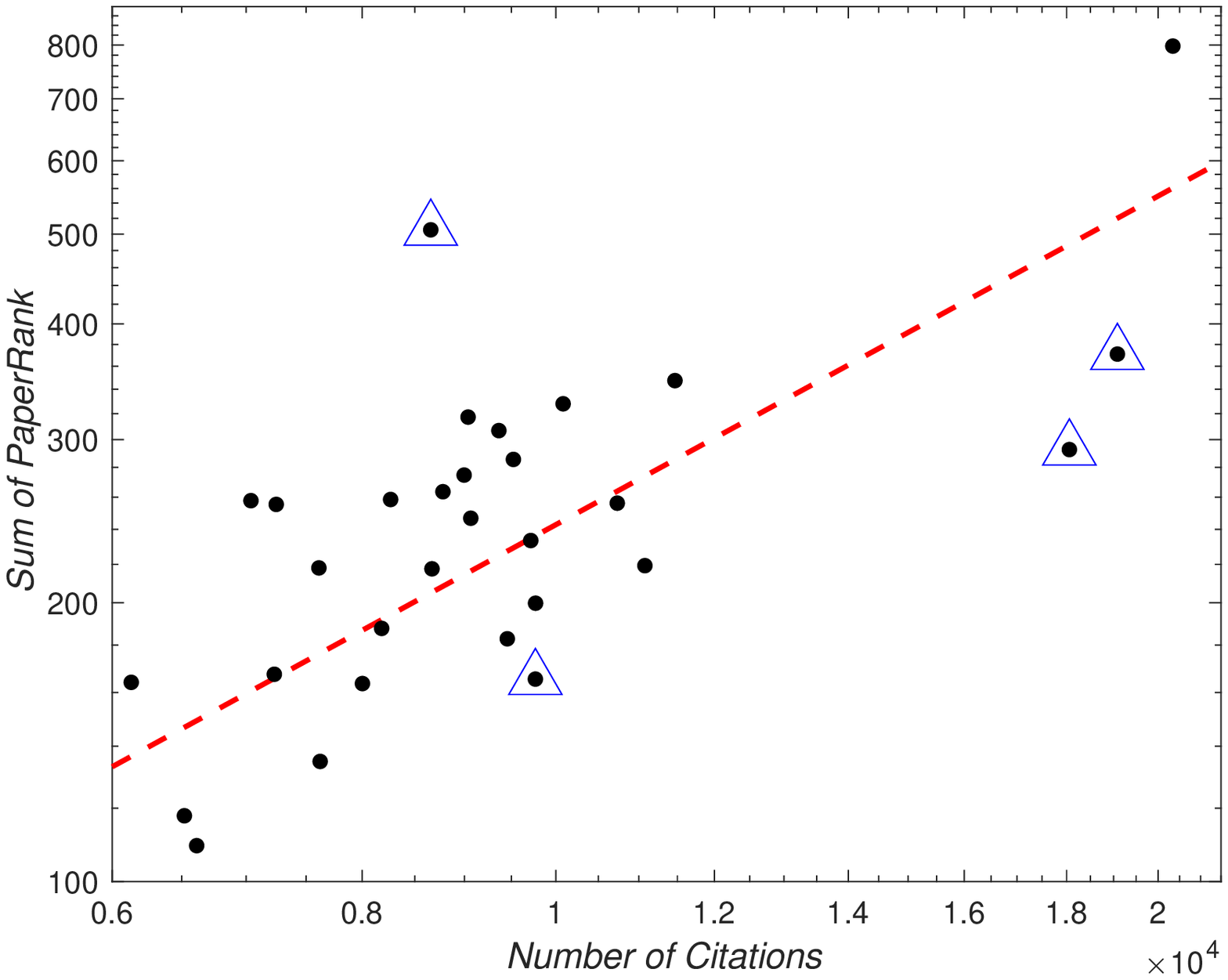}
\vspace*{-1cm}
\end{center}
\caption{Sum of \PapRank\ (\SPR) with respect to the total number of citations (\SC), for the authors with h-index=48 listed in Table\,\ref{Tab1}.} \label{Fig1}
\end{figure}

Analyzing Table \ref{Tab1} we observe that the total number of publication may vary greatly (\NP\ ranges from 68 to 352), as well as the total citations (\SC\ ranges from 6133 to 62815). Moreover, while \NP\ is not significantly related to our parameters, \SC\ seems roughly proportional to the   \SPR\ of the considered author, as is shown in Fig.\,\ref{Fig1}. 
In this figure, \SC\ is on the abscissa, and \SPR\ on the ordinate (for ease of reading the graph, the point with the largest value has been cut, since it is significantly farther from the others). A dashed line indicates the linear regression line between \SC\ and \SPR. 
It can be  immediately seen that many points are close to the regression line but some of them are very far.
In particular, one author (subject area ``Physics and Astronomy'') has a \SPR\ value that is more than double the expected value, while three authors (subject areas ``Environmental Science'' and ``Earth and Planetary Sciences'') have a \SPR\ that is just over half of the expected value: the points corresponding to such authors are marked in Fig.\,\ref{Fig1} with triangles. 

Further, to better understand these results, all the 75 papers having more than 500 citations by the authors listed in Table\,\ref{Tab1} have been analyzed in more details. 
For each paper, we have computed the corresponding ratio
\begin{equation}\label{sig}
\rho = \frac{ \mbox{\NC\ }}{\mbox{{\PapRank\ }}},
\end{equation}
i.e., the average number of citations needed to increase the \PapRank\ of the given paper by one. It turns out that $\rho$ approximately varies in the interval [14,\,67]: 
\begin{itemize}
\item the lowest values  of $\rho$ (say, less than 25) are for papers in ``Physics and Astronomy" and ``Medicine" subject areas;
\item the highest values (greater than 60) are obtained for papers in the ``Environmental Science'' and ``Agricultural and Biological Sciences'' subject areas.  
\end{itemize} 
As is clear, these results are in some agreement with the triangles marked in Fig.\,\ref{Fig1}.

\begin{table}
\begin{center}
\begin{tabular}{rlrrrr}
Scopus\ A.Id \	&	Subj.\ Area	&	\NP	&	\SC	&	$\sum$ P.R.	&	A.R.	\\ \hline
24782201700	&	Agr\&BioSci 	&	151	&	11078	&	219.32	&	28.59	\\
7103056456	&	Bioc\&Ge\&Mo	&	146	&	8003	&	163.60	&	44.98	\\
7006175921	&	Bioc\&Ge\&Mo	&	82	&	9768	&	199.77	&	43.42	\\
7202686127	&	Bioc\&Ge\&Mo	&	142	&	62815	&	2176.80	&	946.37\\
7005804651	&	Chem 	&	218	&	7623	&	134.81	&	44.94	\\
7102471934	&	Chem 	&	170	&	9038	&	317.27	&	108.08	\\
7005780945	&	ComputSci	&	357	&	9066	&	246.75	&	71.42	\\
57201695156	&	Earth\&Plat	&	129	&	9766	&	165.42	&	49.54	\\
7004116785	&	EnvirSci	&	226	&	6133	&	164.09	&	59.94	\\
7004930908	&	EnvirSci	&	122	&	7231	&	167.43	&	48.02	\\
7402266526	&	EnvirSci	&	88	&	18058	&	292.66	&	45.62	\\
7202917464	&	EnvirSci	&	97	&	19084	&	371.04	&	149.06	\\
7006980158	&	Imm\&MicBio	&	99	&	9455	&	182.86	&	32.82	\\
6603637799	&	MaterSci	&	139	&	10731	&	256.17	&	41.40	\\
7006852195	&	Medic	&	193	&	7039	&	257.76	&	62.05	\\
7102809705	&	Medic	&	210	&	7247	&	255.39	&	37.47	\\
7005015688	&	Medic	&	248	&	8182	&	187.65	&	35.70	\\
7202794939	&	Medic	&	315	&	8267	&	258.50	&	58.89	\\
7004446309	&	Medic	&	316	&	8780	&	263.53	&	56.50	\\
6701722543	&	Medic	&	159	&	8997	&	274.68	&	29.04	\\
7004046246	&	Medic	&	319	&	9364	&	306.80	&	24.56	\\
7403276785	&	Medic	&	278	&	9522	&	285.58	&	52.61	\\
7006198419	&	Medic	&	378	&	10083	&	327.92	&	35.61	\\
7101861992	&	Medic	&	163	&	11468	&	347.37	&	35.47	\\
7006253731	&	Medic	&	301	&	20342	&	797.84	&	39.56	\\
55726373700	&	NeuSci	&	68	&	6520	&	117.78	&	40.89	\\
7003644890	&	NeuSci	&	170	&	6613	&	109.35	&	17.79	\\
7103174875	&	Pharm	&	184	&	8670	&	217.59	&	45.95	\\
25641028400	&	Pharm	&	299	&	9714	&	233.45	&	49.72	\\
35228729900	&	Phys\&Astr	&	164	&	7612	&	218.08	&	0.49	\\
7201985737	&	Phys\&Astr	&	352	&	8658	&	505.31	&	119.47	\\
\hline
\end{tabular}
\end{center}
\caption{Sum of \PapRank\ and \AutRank\ for authors with h-index=48 and surname starting with the letters ``AL", along with their number of publications and total citations, grouped by subject area (listed in alphabetical order).}\label{Tab1}
\end{table}

\begin{table}
\begin{center}
\begin{tabular}{ll}
Acronym	&	Subject  Area \\ \hline
Agr\&BioSci 	& Agricultural and Biolog. Sciences \\
Bioc\&Ge\&Mo 	& Bioch., Genetics and Mol. Biology \\
Chem 	& Chemistry \\
ComputSci 	& Computer Science \\
Earth\&Plat 	& Earth and Planetary Sciences \\ 
Engin 	& Engineering \\
EnvirSci 	& Environmental Science \\
Imm\&MicBio 	& Immunology and Microbiology \\
MaterSci 	& Materials Science \\
Math 	& Mathematics \\
Medic 	& Medicine \\
NeuSci 	& Neuroscience \\
Pharm 	& Pharmacol., Toxicol. and Pharmaceut. \\
Phys\&Astr & Physics and Astronomy \\
\hline
\end{tabular}
\end{center}
\caption{Acronyms used by Scopus to denote the considered subject areas.}\label{Tab3}
\end{table}

\begin{figure}[t]
\begin{center}
\vspace*{-1cm}
\includegraphics[width=\textwidth]{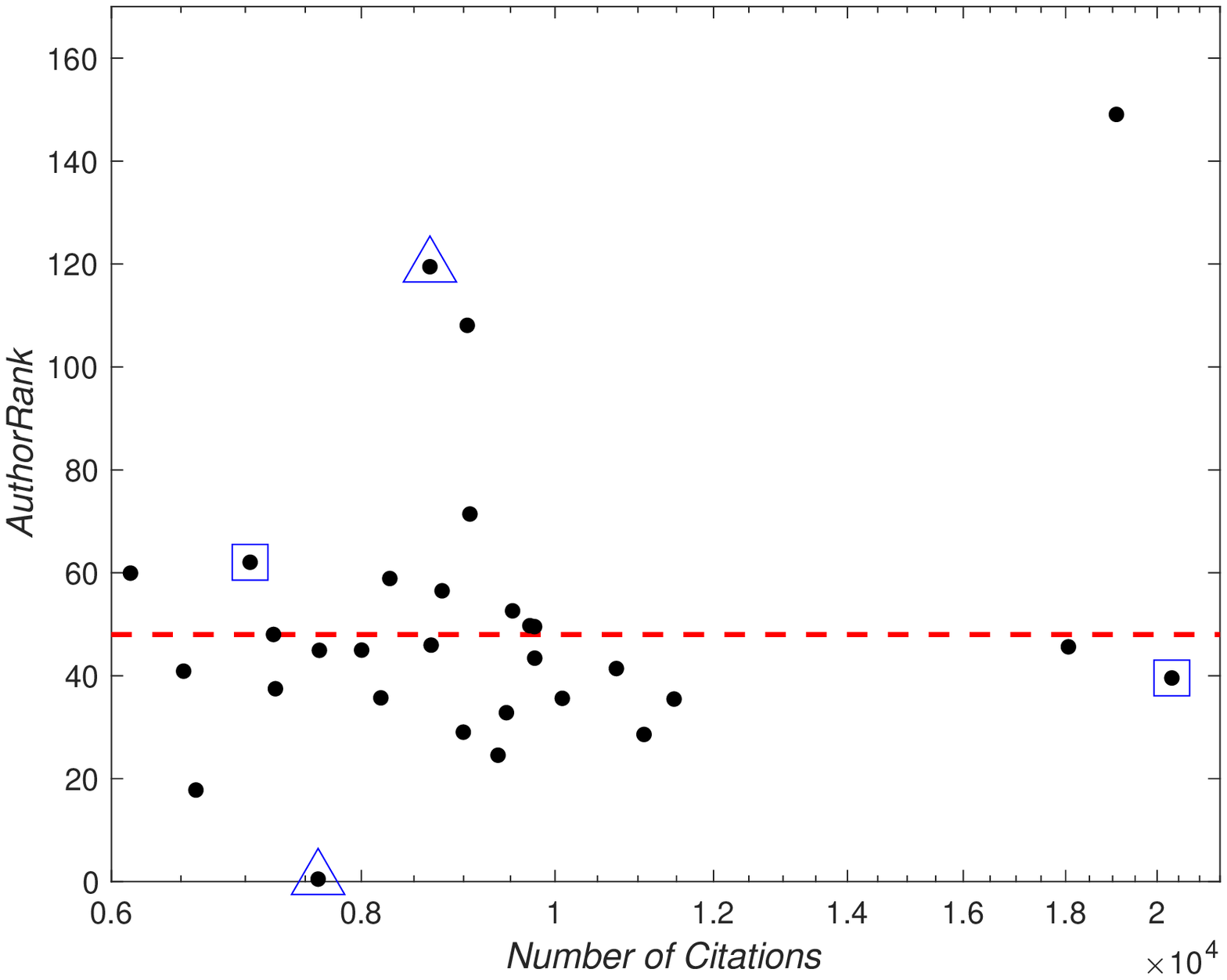}\\
\vspace*{-1cm}
\end{center}
\caption{\AutRank\ with respect to the total number of citations (\SC) for the authors with h-index=48 listed in Table\,\ref{Tab1}.} \label{Fig2}
\end{figure}

The results in Fig.\,\ref{Fig2} are certainly more significant: here the \SC\ value is on the abscissa while the \AutRank\ is on the ordinate (again, the plot does not include the author with the highest  number of citations and \AutRank\ (\AutRank\ $=946.37$), which would appear far, on the upper-right of the figure).
The other authors have an \AutRank\ which varies in a fairly wide range from 0.49 to 149.06. Half of them have \AutRank\ between 35 and 50 (relatively close to the value \AutRank\ $= \text{h-index} = 48$ indicated by the dashed line in the figure), but obviously the authors with values outside this range are meaningful.
In particular, the two authors in the ``Physics and Astronomy'' area listed in Table\,\ref{Tab1} have \AutRank\ equal to 0.49 and 119.47, respectively, despite having a comparable \SC\ number  (see  the triangles in Fig.\,\ref{Fig2}). 
This huge difference depends on the fact that, while the former (an experimental physicist) has publications in collaboration with over 100 authors, the latter (a theoretical physicist) has publications with a few co-authors and the papers citing him have very targeted bibliographies (17 bibliographical items, on average).  

Summing up, even though we have randomly chosen the set of authors, the obtained results allowed to emphasize some remarkable situations: 
\begin{itemize}
\item the author in Table\,\ref{Tab1} with the highest \AutRank\ (Reka Albert, \AutRank$~=946.37$, fourth line in the table) presents in Scopus a publication with over 20 thousand citations (\PapRank$~= 822,92$) and at least 6 others with a number of citations greater than 1000. This value of the \AutRank, comparable as we will see later, with that of Stephen Hawking (\AutRank$~= 984.60$, see Table\,\ref{Tab2}), is not highlighted by the h-index;

\item the 11 authors classified in the Subject class ``Medicine''  have an \AutRank\ ranging from about 24.5 to about 62.  Also for these authors the value \SC\ definitely does not emphasize the obtained ranking. In particular, the author with the highest \AutRank\ has the smallest \SC\ while the author with the highest \SC\ has an \AutRank\ smaller than 40  (see  the squares in Fig.\,\ref{Fig2}).

\end{itemize}

\begin{table}
\hspace{-0.8cm}
\begin{tabular}{rlrrrrrr}
Author Name	&	Subj.\ Area	&	\NP	&	\SC	&	h-ind. & $\sum$P.R.	&	Aut.R.	\\ \hline
Marks II, R.	&	Engin		&	302	&	5674		&	36	&	284.94	&	81.26	\\
Bonner, J.T.	&	Bioc\&Ge\&Mo	&	97	&	5789		&	32	&	185.09	&	96.41	\\
Goodall, J.	&	Agr\&BioSci 	&	59	&	6118		&	35	&	115.04	&	27.28	\\
Higgs, P.		&	Phys\&Astr	&	19	&	6466		&	9	&	182.34	&	182.24	\\
Vestergaard Hau, L.	&	Phys\&Astr.  &	49	&	6847		&	18	&	285.80	&	74.88	\\
Bray, D.		&	Bioc\&Ge\&Mo	&	111	&	10314	&	48	&	268.41	&	118.59	\\
Geller, M.		&	Phys\&Astr 	&	216	&	11993	&	61	&	217.32	&	59.99	\\
Townes, C.H.	&	Phys\&Astr	&	248	&	12670	&	48	&	480.10	&	174.98	\\
Knuth, D.E.	&	Math			&	133	&	12914	&	42	&	701.57	&	488.41	\\
Ogawa, S.		&	Bioc\&Ge\&Mo	&	56	&	14559	&	31	&	413.87	&	91.84	\\
Berners-Lee, T.	&	ComputSci	&	57	&	15424	&	21	&	753.50	&	246.84	\\
Aspect, A. 	&	Phys\&Astr	&	219	&	15519	&	51	&	440.19	&	214.11	\\
Penrose, R.	&	Phys\&Astr	&	133	&	15945	&	45	&	679,65	&	493,46	\\
Watson, J.D.	&	Bioc\&Ge\&Mo	&	128	&	15998	&	39	& 	440.19	&	214.11	\\ 
Wilson, E.	        &	Agr\&BioSci 	&	228	&	20495	&	57	&	481.95	&	288.45	\\
Montagnier, L.	&	Medic		&	380	&	24647	&	71	&	760.46	&	136.71	\\ \hdashline
Hawking, S.	&	Phys\&Astr	&	163	&	41757  	&	75	&	1339.02	&	984.60 \\
Karplus, M.	&	Bioc\&Ge\&Mo	&	812	&	119539	& 	157	& 	2754.81	&	1010.29	\\
\hline
\end{tabular}
\caption{Sum of \PapRank\ (\SPR) and \AutRank\ (Aut.R.) for the authors in \cite{BestSchool} with \SC\ in the range [5000,\,25000], sorted by increasing number of citations.} \label{Tab2}
\end{table}

\begin{figure}[t]
\begin{center}
\vspace*{-2mm}
\includegraphics[width=\textwidth]{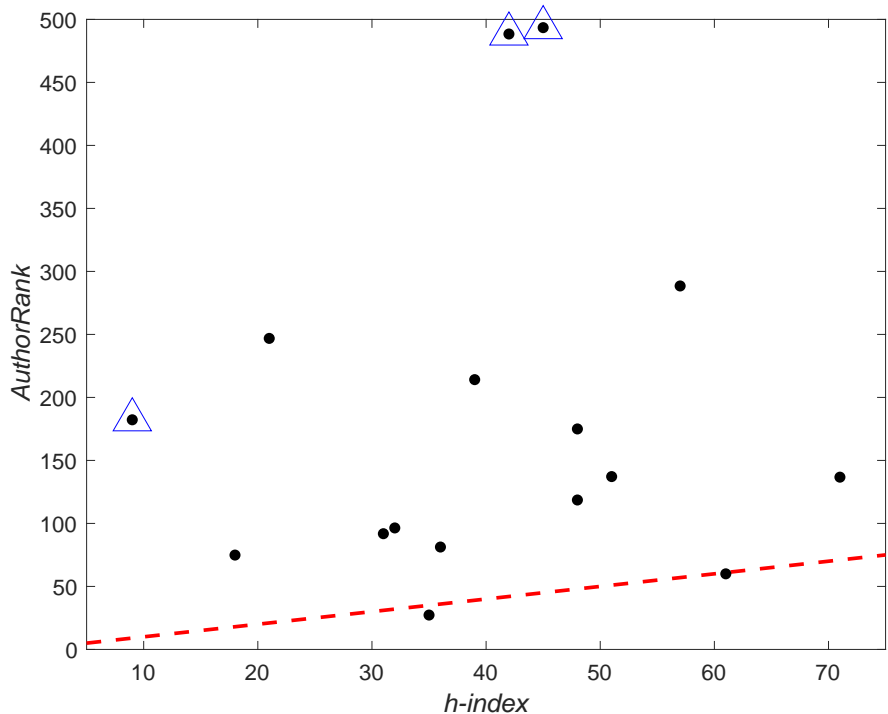}
\vspace*{-1cm}
\end{center}
\caption{\AutRank\ with respect to h-index for the authors in Table\,\ref{Tab2}.}\label{Fig3}
\end{figure}

Fig.\,\ref{Fig2} also shows that more than 50\% of the authors have an \AutRank\ smaller than the h-index (in particular, two of them,  corresponding to the rightmost dots  below the dashed-line, having  a very high number of citations). We now further investigate this aspect, by considering some of the 50 most influential scientists in the world today \cite{BestSchool}: they are listed in Table\,\ref{Tab2}, along with the previously mentioned  authors Hawking and Karplus. In particular, we consider the authors in \cite{BestSchool} with \SC\ ranging from 5000 to 25000, according to the Scopus database, 
resulting into the 16 scientists listed in Table\,\ref{Tab2}. In Fig.\,\ref{Fig3}, we plot the \AutRank\ of each author w.r.t the corresponding h-index. From the figure, one may see that for almost all these authors, the \AutRank\ is much greater than the corresponding h-index (see the dashed line representing $\AutRank=$ h-index): almost 20 times for Higgs who has h-index $= 9$, more than 10 times for Knuth and Penrose who have  \AutRank\ $\approx 500$, which are marked with the triangles in Fig.\,\ref{Fig3}).

Incidentally, the most cited papers of Knuth and Wilson have approximatively the same \PapRank\ (179.63 and 175.12, respectively), though having a quite different number of citations (approximately 3800 and 8600, respectively). This means that also papers from authors in different subject areas are more effectively compared.

\section{New indices and future work} \label{SecNewind}

Although the results obtained in the previous section seem very encouraging, possible requests of different indicators from some subject areas may require a non-standard use of \PapRank\ and \AutRank. In fact, the simplicity of their definition could give a simple answer to these questions, too. For example, for some subject areas, where it is usual to have papers with many authors, the partitioning of the \PapRank\ among authors could be proportional to the order of authors, rather than being uniform, as sketched in Remark~\ref{dispro}.

Furthermore, to obtain indices that also take into account both the number of papers and their quality, it could be possible to reformulate h-index, h$^{\text{c}}$-index, and i10-index by replacing,  in the original definition, the number of citations of each paper with the ratio between \PapRank\ and the number of authors. For example, we could define the following new indices, for an author $A_k$, whose \AutRank\ is given by
(\ref{Arank}), which correspond to the h-index and i10-index, respectively:
\begin{itemize}
\item h$_{\alpha}$-index: the largest value $p$ s.t.\, $\displaystyle \sum_{j=1}^{m_k} \left(\frac{\PapRank(P_{b_{j,k}})}{\NumAut(P_{b_{j,k}})} \ge \alpha p \right) \ge p$,
\item i$_{\beta}$-index $= \displaystyle \sum_{j=1}^{m_k} \left(\frac{\PapRank(P_{b_{j,k}})}{\NumAut(P_{b_{j,k}})} \ge \beta \right)$,
\end{itemize}
where $\alpha$ and $\beta$ are positive quantities suitably chosen.
In Table\,\ref{Tab4} we list some examples for such indices, for the authors  with h-index=48 listed in Table\,\ref{Tab1},  by choosing $\alpha=0.01$ and $\beta=0.1$, and compare them with the usual h-index and i20 index, respectively.\footnote{Indeed, we found the latter index more significant than i10.} We observe that the ``outlier'' author not shown in Figs.~\ref{Fig1} and \ref{Fig2}  (A.Id 7202686127) exhibits values similar to most of the other authors.  Moreover, the difference between the two authors in Fig.\,\ref{Fig2} in the ``Physics and Astronomy'' subject area is still very evident. This means that the combination of \AutRank\ and one of these new parameters may provide a more comprehensive information on a given author.

\begin{table}
\begin{center}
\begin{tabular}{rrrrrrc}
Scopus\ A.Id\	&	A.R.	& h$_{.01}$ & h & i$_{.1}$ & i20 \\ \hline
7202686127	& 946.37	& 41	& 48	& 84	& 81 & outlier \\
57201695156	& 49.54	& 37	& 48	& 64	& 83  & left-lower $\triangle$ in Fig.\,\ref{Fig1} \\
7202917464	& 149.06	& 39	& 48	& 71	& 75  & right-lower $\triangle$ in Fig.\,\ref{Fig1} \\
7402266526	& 45.62	& 26	& 48	& 55	& 66 & middle-lower $\triangle$ in Fig.\,\ref{Fig1} \\
7006253731	& 39.56	& 30	& 48	& 69	& 72 & right $\square$ in Fig.\,\ref{Fig2} \\
7006852195	& 62.05	& 38	& 48	& 94	& 90 & left $\square$ in Fig.\,\ref{Fig2} \\
7201985737	& 119.47	& 52	& 48	& 145 & 117 & upper $\triangle$ in Figs.\,\ref{Fig1} and \ref{Fig2} \\
35228729900	& 0.49	& 2	& 48	& 0	& 104 & lower $\triangle$ in Fig.\,\ref{Fig2} \\
\hline
\end{tabular}
\end{center}
\caption{Indices for some authors in Table\,\ref{Tab1}.}\label{Tab4}
\end{table}

Obviously, by considering only the most recent papers, one may derive an analog of the h$^{\text{c}}$-index (see also \cite{AmBr14}).
 
Last but not least, it is worth mentioning that one could, in principle, evaluate journals or research groups (or Universities) by combining the associated \PapRank\ and \AutRank: this could represent a subsequent goal of this research.

\section{Conclusions} \label{SecLast}
In this paper, we show the result of a preliminary study on the application of the \PapRank\ and \AutRank\ indices proposed in \cite{AmBr14} to the Scopus database. The two indices give more interesting values than the bare number of citations and h-index, respectively, with practically the same computational cost. This fact allows to easily keep them always updated. They can also be used for the definition of new indices and to evaluate, always quantitatively, scientific journals and research groups.

\section*{Acknowledgement} We are very grateful to Elsevier, in particular to the Integration Support of Scopus, for the assistance and for granting us the intensive use of the API keys needed for this research. 
We are also grateful to the reviewers, for their comments and remarks.



\begin{thebibliography}{99}

\bibitem{AmBr14} 
Amodio, P., \& Brugnano, L. (2014). Recent advances in bibliometric indexes and the PaperRank problem, {\it Journal of Computational and Applied Mathematics}, 267, 182--194.

\bibitem{Be07} 
Bergstrom, C.\,T. (2007). 
Eigenfactor, {\it College \& Research Libraries News}, 68(5), 314--316.

\bibitem{BoRoVa06} 
Bollen, J., Rodriguez, M.\,A., \& van de Sompel, H. (2006). Journal status. {\it Scientometrics}, 69(3), 669--687.

\bibitem{BVdSHC09} 
Bollen, J., van de Sompel, H., Hagberg, A., \& Chute, R. (2009). A principal component analysis of 39 scientific impact measures. {\it PLoS ONE}, 4(6), e6022.

\bibitem{BoMuDa08} 
Bornmann, L., Mutz, R., \& Daniel, H.-D. (2008).
Are there better indices for evaluation purposes than the h-index? A comparison of nine different variants of the h index using data from biomedicine. {\it Journal of the American Society for Information Science and Technology}, 59(5), 830--837.

\bibitem{BrPa12} 
Brin, S., \&  Page L. (2012). Reprint of: The Anatomy of a Large-Scale Hypertextual Web Search Engine. {\it Computer Networks}, 56(18), 3825--3833.

\bibitem{BrNa21}
Brito, R., \& Navarro A.\,R. (2021). The inconsistency of h-index: A mathematical analysis. {\it Journal of Informetrics},
15(1), 101--106.

\bibitem{ChXiMaRe06}
Chen, P., Xie, H., Maslov, S., \& Redner, S. (2006).
Finding scientific gems with Google’s PageRank algorithm.
{\it Journal of Informetrics}, 1(1), 8--15.

\bibitem{DoMe15}
Dorogovtsev, S.\,N., \& Mendes J.\,F.\,F. (2015). Ranking scientists. {\it Nature Physics}, 11(11), 882--883.

\bibitem{Eg06} 
Egghe, L. (2006). An improvement of the h-index: the g-index. {\it ISSI Newsletter}, 2(1), 8--9.

\bibitem{Eg06b} 
Egghe, L. (2006). Theory and practise of the g-index. {\it Scientometrics}, 69(1), 131--152.

\bibitem{Fiala} Fiala, D. (2012). Time-aware PageRank for bibliographic networks. {\it Journal of Informetrics}, 6(3), 370--388.

\bibitem{GaHiAlFa18} 
Garner, R.\,M., Hirsch, J.\,A., Albuquerque, F.\,C., \& Fargen, K.\,M. (2018). Bibliometric indices: defining academic productivity and citation rates of researchers, departments and journals. {\it Journal of NeuroInterventional Surgery}, 10(2), 102--106.

\bibitem{Hi05} 
Hirsch, J.\,E. (2005). An index to quantify an individuals scientific research output. {\it Proceedings of the National Academy of Sciences of the United States of America}, 102(46), 16569--16572.

\bibitem{KaHaGo04} 
Kamvar, S., Haveliwala, T., \& Golub, G.\,H. (2004).
Adaptive methods for the computation of PageRank.
{\it Linear Algebra and its Applications}, 386(1-3 suppl.), 51--65.

\bibitem{KaHaMaGo03} 
Kamvar, S.\,D., Haveliwala, T.\,H., Manning, C.\,D., \& Golub, G.\,H. (2003).
Extrapolation methods for accelerating PageRank computations.
{\it Proceedings of the 12th International Conference on World Wide Web, WWW 2003}, 261-270.

\bibitem{KrKoNg21}
Kreinovich, V., Kosheleva, O., \& Nguyen H.\,P. (2021). Why h-Index. {\it Studies in Computational Intelligence}, 899, 61--65.

\bibitem{LW09} 
Li, J.\, Willett P. (2009). ArticleRank: a PageRank-based alternative to numbers of citations for
analysing citation networks. {\it Aslib Proceedings: New Information Perspectives}, 61(6), 605--618.

\bibitem{LiBoNeVa05} 
Liu, X.\, Bollen, J.\, Nelson, M.\,L., \& Van De Sompel H. (2005).
Co-authorship networks in the digital library research community.
{\it Information Processing and Management}, 41(6), 1462--1480.

\bibitem{MaGuZh07} 
Ma, N., Guan, J., \& Zhao, Y. (2007)
Bringing PageRank to the citation analysis
{\it Information Processing \& Management}, 44(2), 800-810.

\bibitem{MoNyMoAnSa21}
Mohammed, S., Nyantakyi, E.\,K., Morgan, A., Anumah, P., \& Sarkodie-kyeremeh, J. (2021). Use of relative extra citation counts and uncited publications to enhance the discriminatory power of the h-index. {\it Scientometrics},
126(1), 181--199.

\bibitem{PaBr98} 
Page, L., \& Brin, S. (1998). The Anatomy of a Large-Scale Hypertextual Web Search Engine. {\it Computer Networks}, 30(1-7),  107--117.

\bibitem{SiKaMa07} 
Sidiropoulos, A., Katsaros, D.\, \& Manolopoulos, Y. (2007). Generalized Hirsch h-index for disclosing latent facts in citation networks. {\it Scientometrics}, 72, 253--280.

\bibitem{WaXiYaMa07} 
Walker, D.\, Xie, H., Yan, K.\,K., \& Maslov S. (2007). Ranking scientific publications using a model of network traffic. {\it Journal of Statistical Mechanics: Theory and Experiment}, 6, P06010.

\bibitem{BeWe10} 
West, J.\,D., Bergstrom, T.\,C., \& Bergstrom, C.\,T. (2010). The eigenfactor metrics$^{\text{TM}}$: A network approach to assessing scholarly journals.  {\it College \& Research Libraries News}, 71(3), 236--244.

\bibitem{YaDiSu11} 
Yan, E., Ding, Y., \& Sugimoto, C.\,R. (2011).
P-Rank: An indicator measuring prestige in heterogeneous scholarly networks.
{\it Journal of the American Society for Information Science and Technology}, 62(3), 467--477.

\bibitem{Zh09} 
Zhang, C.\,T. (2009). The e-index, complementing the h-index for excess citations. {\it PLoS One}, 4(5), e5429.

\bibitem{Google} {\it Google} web page: \url{http://www.google.it/}

\bibitem{GScholar} {\it Google Scholar} web page: \\ \url{https://scholar.google.com/ intl/en/scholar/about.html}

\bibitem{Scopus} {\it Scopus} web page: \url{https://www.scopus.com/}

\bibitem{apik} \url{https://dev.elsevier.com/sc_apis.html}

\bibitem{BestSchool} 
{\it The 50 most influential Scientists in the world} web page: \\
\url{https://thebestschools.org/features/50-influential} \url{-scientists-world-today/}

\end{thebibliography}
\end{document}